\documentclass[preprint,12pt]{elsarticle}
\usepackage[nodots,nocompress]{numcompress}
\usepackage{amsmath}

\begin{document}
\begin{frontmatter}
\title{Reciprocal link for a three-component Camassa-Holm type equation}

\author{Nianhua Li}

\cortext[cor]{Corresponding author. linianh@hqu.edu.cn}
\address{School of Mathematics,
Huaqiao University,
Quanzhou, Fujian 362021, People's Republic of China}

\begin{abstract}
we discuss a reciprocal transformation for a three-component Camassa-Holm type equation and find that the transformed system is a reduction of the first negative flow for an extended MKdV hierarchy.
\bigskip

\end{abstract}

\end{frontmatter}
\section{Introduction}
The Camassa-Holm (CH) equation
\begin{equation}\label{ch}
u_t-u_{xxt}+3uu_x=2u_xu_{xx}+uu_{xxx},
\end{equation} was derived by Camassa and Holm as a model for unidirectional dispersive shallow-water motion using an asymptotic expansion directly in the Hamiltonian for Euler's equations\cite{Holm,Hyman}. It is completely integrable in the sense of Lax pair, bi-Hamiltonian structure, infinitely many conserved quantities etc. The equation can be solved by the inverse scattering transformation\cite{Con1}, and is found to admit multi-solitons solutions, algebro-geomrtric solutions and so on\cite{Liya,Mat,Gez}. Especially, the discovery of peakons make the CH equation the subject of extensive research in recent years\cite{Holm,Hyman}. The peakons is a weak solution in some Sobolov spaces and is interesting to research in water wave theory and in the view of general analysis for PDEs. Besides, the CH equation has a reciprocal link to the first negative flow of the KdV hierarchy\cite{fuch}, and the conservation laws of each may be connected\cite{lene}. Subsequently, many other CH type equations (the equations possess peakon solutions) are proposed and studied, such as the DP equation, the Novikov's equation and the Geng-xue equation\cite{Deg,gas,Novikov,Geng,li3,hon3,wang}. Although many of them is integrable but it has some nonstandard features such as the weak Painlev\'e property,  and is still much to understand via reciprocal transformation\cite{hon2}. 

Recently, a three-component CH type system admitting the following 3$\times$ 3 spectral problem
\begin{equation}\label{thr1}
\psi_x=\left(
         \begin{array}{ccc}
           0 & 1 & 0 \\
           1+\lambda v &0 & u \\
           \lambda w & 0 & 0 \\
         \end{array}
       \right)\psi.
\end{equation} is proposed by Geng and Xue \cite{Geng2} with $N$-peakon solutions
\begin{eqnarray}
&&u_{t}=-v p_{x}+u_{x}q+\frac{3}{2}u q_{x}-\frac{3}{2} u(p_{x}r_{x}-pr),\nonumber\\
&&v_{t}=2vq_{x}+v_{x}q ,\label{h1}\\
&&w_{t}=v r_{x}+w_{x} q+\frac{3}{2}w q_{x}+\frac{3}{2} w(p_{x}r_{x}-pr),\nonumber
\end{eqnarray}
where
\begin{eqnarray*}
&&u=p-p_{xx},\nonumber\\
&&v=\frac{1}{2}(q_{xx}-4q+p_{xx}r_{x}-r_{xx}p_{x}+3p_{x}r-3p r_{x}),\\
&&w=r_{xx}-r.\nonumber
\end{eqnarray*}  The bi-Hamiltonian structure as well as infinite many conserved quantities of this three-component CH type equation and the dynamical system for the $N$-peakon solutions of it are also obtained\cite{Geng2,Li}. 

Very recently, another three-component CH type system \cite{popo} associated a 3$\times$ 3 spectral problem
\begin{equation}\label{thr2}
\phi_{x}=\left(
    \begin{array}{ccc}
      0 & 0 & 1 \\
       \lambda m_{1}& 0 & \lambda m_{3}\\
      1 &  \lambda m_{2}& 0 \\
    \end{array}
  \right)\phi.
\end{equation}
 is constructed by us
\begin{eqnarray}
&&m_{1t}+u_2gm_{1x}-m_{3}(u_{2x}f-u_2g)-m_1(3u_2f-m_3u_2)=0,\nonumber\\
&&m_{2t}+u_2gm_{2x}+m_{2}(3u_{2x}g+m_3u_2)=0, \label{eqq}\\
&&m_{3t}+u_2gm_{3x}-m_3(2u_2f+u_{2x}g-m_3u_2)=0, \nonumber
\end{eqnarray}
where
\begin{equation*}
m_{i}=u_{i}-u_{ixx}, \ i=1,2,3, \quad f=u_{3}-u_{1x},\quad g=u_{1}-u_{3x}.
\end{equation*}However, a detail calculation shows the spectral (\ref{thr2}) can be rewritten as (\ref{thr1}), so above two systems can be contained in a same hierarchy. Bi-Hamiltonian structure and infinitely many conserved quantities for the CH type system (\ref{eqq}) are worked out.

The propose of this paper is to construct a reciprocal transformation for the three-component CH type equation(\ref{eqq}), and the relationship between the transformed system and an extended MKdV hierarchy is obtained. This will be done in Sec. 2 and Sec. 3.
\section{A reciprocal transformation}
In this section, we will construct a reciprocal transformation for the CH type system (\ref{eqq}). As pointed out in Ref. \cite{popo}, the three-component system (\ref{eqq}) is the compatible condition of the spectral problem (\ref{thr2}) and associated  auxiliary problem
\begin{equation}\label{lat}
\phi_t=\left(
         \begin{array}{ccc}
           -u_2f & \lambda^{-1}u_2 & -u_2g \\
          \lambda^{-1}f-\lambda m_1u_2g & u_2f+u_{2x}g-\lambda^{-2} & \lambda^{-1}g-\lambda m_3u_2g \\
           -u_{2x}f & \lambda^{-1}u_{2x}-\lambda m_2u_2g & -u_{2x}g \\
         \end{array}
       \right)
\phi.
\end{equation}Based on above Lax pair, an infinite sequence of conservation laws may be constructed.
Especially, one of them is
\begin{equation*}
((m_2m_3)^{\frac{1}{2}})_t=-((m_2m_3)^{\frac{1}{2}}u_2g)_x,
\end{equation*}which defines a reciprocal transformation
\begin{equation}\label{trans}
dy=adx-au_2gdt,\quad d\tau=dt,
\end{equation}where $a=(m_2m_3)^{\frac{1}{2}}$.

In the following, we will study the transformed system of (\ref{eqq}). On the one hand,
writing the column vector $\phi$ as $\phi=(\phi_1,\phi_2,\phi_3)^{T}$ and elimating $\phi_3$ from the spectral problem (\ref{thr2}), we get
\begin{eqnarray*}
&&\phi_{1xx}-\phi_1=\lambda m_2\phi_2, \quad \phi_{2x}=\lambda m_1\phi_1+\lambda m_3\phi_{1x}.
\end{eqnarray*}which in the new variable can be written as
\begin{eqnarray}
&&\phi_{1yy}+\frac{a_y}{a}\phi_{1y}-\frac{1}{a^2}\phi_1=\lambda \frac{1}{m_3}\phi_2, \label{lax1}\\
&&\phi_{2y}=\lambda \frac{m_1}{a}\phi_1+\lambda m_3\phi_{1y}.\label{lax2}
\end{eqnarray} Setting $b=e^{-\partial_y^{-1}\frac{m_1}{am_3}}$, by making a gauge transformation
\begin{equation*}
\phi_1=b\varphi_1, \quad \phi_2=m_3b\varphi_2,
\end{equation*}
the spectral problem (\ref{lax1})-(\ref{lax2}) is transformed to
\begin{eqnarray}\label{q1}
&&\varphi_{1yy}-Q_2\varphi_{1y}-Q_1\varphi_1=\lambda \varphi_2,\\\label{q2}
&& \varphi_{2y}=\lambda\varphi_{1y}+Q_3\varphi_2,
\end{eqnarray}
where
\begin{eqnarray*}
&&Q_1=-\frac{b_{yy}}{b}-\frac{a_yb_y}{ab}+\frac{1}{a^2}, \quad Q_2=-2\frac{b_y}{b}-\frac{a_y}{a},\quad Q_3 =\frac{m_1}{am_3}-\frac{m_{3y}}{m_3}.
\end{eqnarray*}

It is easy to find that the scalar form of the spectral problem (\ref{q1})-(\ref{q2}) is
\begin{equation}\label{exten}
\partial_y^2+u\partial_y+v+\partial_y^{-1}w,
\end{equation} which  is the spectral problem of an extended modified KdV hierarchy\cite{Oevel},  and this extended MKdV hierarchy is related to 
the extended KdV hierarchy(the Yajima-Oikawa hierarchy\cite{Cheng,Yaji}) via a Miura transformation\cite{Oevel}. Furthermore, as we already mentioned\cite{li3}, the transformation between $(Q_1,Q_2,Q_3)$ and $(u,v,w)$ may be also connected via a Miura transformation. Indeed, the Miura transformation
\begin{equation}\label{miu}
u=-(Q_2+Q_3), \quad v=Q_2Q_3-Q_1+Q_{3y}, \quad w=Q_1Q_3-(Q_2Q_3)_y-Q_{3yy}
\end{equation}
is obtained from factorizing the Lax operator (\ref{exten}), that is
\begin{equation*}
\partial_y^2+u\partial_y+v+\partial_y^{-1}w=\partial_y^{-1}(\partial_y-Q_3)(\partial_y^2-Q_2\partial_y-Q_1).
\end{equation*}Moreover,  the spectral problem (\ref{q1})-(\ref{q2}) is reduced to the spectral problem of the MKdV hierarchy as $Q_1=Q_2=0$, and we called the hierarchy associated with this spectral problem an extend MKdV hierarchy too.

Similarly, the auxiliary problem (\ref{lat}) may be changed to the following form
\begin{eqnarray*}
&&\varphi_{1\tau}=\lambda^{-1}q_3\varphi_2,\\
&& \varphi_{2\tau}=(-\lambda^{-2}+q_3)\varphi_2+\lambda^{-1}q_2\varphi_{1y}+\lambda^{-1}q_1\varphi_1,
\end{eqnarray*}
where
\begin{eqnarray*}
&&q_1=\frac{u_3-u_{1y}a}{m_3}-\frac{m_1}{m_3^2}(u_1-u_{3y}a), \quad q_2=\frac{a}{m_3}(u_1-u_{3y}a), \quad q_3=m_3u_2.
\end{eqnarray*}
Then the Lax pair for the three-component CH type system (\ref{eqq}) in the new variable may be converted into
\begin{equation}\label{miulax}
\varphi_y=\left(
                  \begin{array}{ccc}
                    0 & 1 & 0 \\
                    Q_1 & Q_2 & \lambda \\
                    0 & \lambda & Q_3 \\
                  \end{array}
                \right)
  \varphi,
\end{equation}
and
\begin{equation}\label{trelax}
\varphi_\tau =\left(
                   \begin{array}{ccc}
                     0 & 0 & \lambda^{-1}q_3 \\
                     0 & q_3 & \lambda^{-1}(q_{3y}+Q_3q_3) \\
                     \lambda^{-1}q_1 & \lambda^{-1}q_2 & -\lambda^{-2}+q_3 \\
                   \end{array}
                 \right)
\varphi.
\end{equation}

On the other hand, we can now to state the transformed system for (\ref{eqq}). Under the transformation (\ref{trans}), the CH type system (\ref{eqq}) is changed to
\begin{eqnarray*}
&&m_{1\tau}=m_1(3u_2f-m_3u_2)+m_3(u_{2y}af-u_2g),\quad  m_1=u_1-a(u_{1y}a)_y,\\
&&m_{2\tau}=-m_2(3u_{2y}ag+m_3u_2),  \quad \quad \quad \quad \quad \quad \quad \quad  m_2=u_2-a(u_{2y}a)_y,\\
&&m_{3\tau}=m_3(2u_2f+u_{2y}ag-m_3u_2), \quad \quad \quad \quad \quad \quad  m_3=u_3-a(u_{3y}a)_y.
\end{eqnarray*}
Through tedious but direct calculations, we get the system for $Q_1, Q_2, Q_3$ under the Liouville  transformation
\begin{equation*}
\left(
  \begin{array}{c}
    Q_1 \\
    Q_2 \\
    Q_3 \\
  \end{array}
\right)=\left(
            \begin{array}{c}
             (m_2m_3)^{-1}((\frac{m_1}{m_3})_x+1-(\frac{m_1}{m_3})^2)\\
             (m_2m_3)^{-\frac{3}{2}}(2m_1m_2-\frac{1}{2}(m_2m_{3})_x) \\
             m_2^{-\frac{1}{2}}m_3^{-\frac{3}{2}}(m_1-m_{3x})\\
            \end{array}
          \right), \quad y=\int_{-\infty}^{x} (m_2m_3)^{\frac{1}{2}} dx,
\end{equation*}that is
\begin{eqnarray}\label{treq}
&&Q_{1\tau}=Q_1q_3-q_1, \quad \quad \quad \quad  \quad S_1=((\partial+Q_3-Q_2)(\partial+Q_3)-Q_1)q_3=-1,\nonumber\\
&&Q_{2\tau}=2q_{3y}+Q_3q_3-q_2, \quad \quad S_2=(-\partial+Q_3)q_1-Q_1q_2=0, \\
&&Q_{3\tau}=-Q_3q_3+q_2, \quad \quad \quad \quad  \  S_3=(-\partial+Q_3-Q_2)q_2-q_1=-1.\nonumber
\end{eqnarray}

Direct calculation shows that the compatible condition of the Lax pair (\ref{trelax}) is just the transformed system (\ref{treq}), therefore under the reciprocal transformation (\ref{trans}), the three-component CH type system(\ref{eqq}) and its Lax pair are changed to  which of (\ref{treq}) accordingly.

\section{The relation between the transformed CH type system and an extended MKdV hierarchy}
According to Ref. \cite{Oevel}, the extended MKdV hierarchy associated with the spectral problem (\ref{exten}) may be formulated as a bi-Hamiltonian system
admitting the following Hamiltonian pair
\begin{eqnarray*}
  {\cal J}_1 &=& \left(
                   \begin{array}{ccc}
                     0 & 0 & 2\partial_y \\
                     0 & 2\partial_y & \partial_y^2+u\partial_y \\
                     2\partial_y & -\partial_y^2+\partial_y u & 0 \\
                   \end{array}
                 \right),\\
  {\cal J}_2 &=& \left(
                   \begin{array}{ccc}
                     6\partial_y & * & * \\
                     4u\partial_y & 2\partial_y^3+2u\partial_y u+\partial_y v+v\partial_y & * \\
                    2\partial_y^3-2\partial_y u\partial_y+2v\partial_y & \theta_1 & \theta_2 \\
                   \end{array}
                 \right),
\end{eqnarray*}where
\begin{eqnarray*}
&&\theta_1=-\partial_y^4+\partial_y^3 u+\partial_y u\partial_y^2-\partial_y u\partial_y u-v\partial_y^2+v\partial_y u+2w\partial_y+\partial_y w,\\
&& \theta_2=\partial_y uw+uw\partial_y+w\partial_y^2-\partial_y^2 w.
\end{eqnarray*}
Based on the Miura link (\ref{miu}), the compatible Hamiltonian operators for another extend MKdV hierarchy of $Q=(Q_1,Q_2,Q_3)^{T}$ may be obtained, which are
\begin{equation*}
\tilde{{\cal J}}_i={\cal F}^{-1}{\cal J}_i{\cal F}^{-1*},  \ i=1, 2,\quad \quad {\cal F}=\left(
    \begin{array}{ccc}
      0 & -1 & -1 \\
      -1 & Q_3 & \partial_y+Q_2 \\
      Q_3 & -\partial_y Q_3 & Q_1-\partial_y Q_2-\partial_y^2 \\
    \end{array}
  \right).
\end{equation*}Then the first negative flow for this extend MKdV hierarchy is obtained, that is
\begin{equation}\label{firstflow}
\left(
  \begin{array}{c}
    Q_1 \\
    Q_2 \\
    Q_3 \\
  \end{array}
\right)_\tau={\cal F}^{-1}{\cal J}_1\left(
                              \begin{array}{c}
                               A \\
                               B \\
                               C\\
                              \end{array}
                            \right), \quad \quad \quad {\cal F}^{-1}{\cal J}_2\left(
                              \begin{array}{c}
                               A \\
                               B \\
                               C\\
                              \end{array}
                            \right)=0.
\end{equation}

In ordering to get the relationship between the transformed system (\ref{treq}) and the first negative flow of the extend MKdV hierarchy for $Q$, we introduce
\begin{eqnarray}\label{abc}
  A &=& \frac{1}{4}(Q_2+Q_3)\partial_y^{-1}(S_1-S_3)-\frac{1}{2}\partial_y^{-1}S_2+\frac{1}{4}(S_1+S_3)-Q_3q_{3y}-Q_3^2q_3,  \nonumber\\
  B &=& \frac{1}{2}\partial_y^{-1}(S_1-S_3)-Q_3q_3, \\
  C &=&-q_3,\nonumber
\end{eqnarray}(here all integration constants are assumed to be zero). Substituting (\ref{abc}) into (\ref{firstflow}), we arrive at the following system
for the first negative flow
\begin{equation}
\left(
  \begin{array}{c}
    Q_1 \\
    Q_2 \\
    Q_3 \\
  \end{array}
\right)_\tau=\left(
           \begin{array}{c}
           Q_1q_3-q_1 \\
           2q_{3y}+Q_3q_3-q_2 \\
           -Q_3q_3+q_2\\
           \end{array}
           \right),\quad {\cal F}^{-1}{\cal K} \left(
                                    \begin{array}{c}
                                      S_1 \\
                                      S_2 \\
                                      S_3 \\
                                    \end{array}
                                  \right)=0,
\end{equation}where
\begin{equation*}
{\cal K}=\left(
               \begin{array}{ccc}
                \frac{1}{2}\partial_y u\partial_y^{-1}-\frac{1}{2}\partial_y & -3 & \frac{3}{2}\partial_y-\frac{1}{2}\partial_y u\partial_y^{-1} \\
                 Q_3\partial_y+v+\frac{1}{2}v_y\partial_y^{-1} & -2u & u\partial_y-\partial_y^2-v-\frac{1}{2}v_y\partial_y^{-1} \\
                 \chi_1 & \partial_y u-\partial_y^2-v & \chi_2 \\
               \end{array}
             \right),
\end{equation*}herein
\begin{eqnarray*}
&&\chi_1=-Q_3\partial_y^2-(2Q_{3y}+Q_2Q_3)\partial_y+\frac{1}{2}(3w+w_y\partial_y^{-1}),\\
&&\chi_2=\partial_y^3+v\partial_y-\partial_y u\partial_y-\frac{1}{2}(3w+w_y\partial_y^{-1}).
\end{eqnarray*}and $(u,v,w)$ are given by (\ref{miu}).

Noticing that the operator $\partial_y^{-1}$ comes from the definition of $A, B, C$, then if $S_1=S_3=-1$ and $S_2=0$, we have ${\cal K}(S_1,S_2,S_3)^{T}=0$, so the transformed three-component CH type system (\ref{treq}) is a reduction of the first negative flow for the extended MkdV hierarchy possessing the spectral problem (\ref{miulax}).

{\bf Remark}. The relationship of Hamiltonian structures between the three-component CH type system (\ref{eqq}) and the transformed system (\ref{treq}) may be obtained following the step in Ref. \cite{lic} with the help of the Liouville transformation.

\bigskip
\noindent
{\bf Acknowledgments}

This work is partially supported by the National Natural Science Foundation of China (Grant Nos. 11401572 and 11401230) and the Initial Founding of Scientific Research for the introduction of talents of Huaqiao University (Project No. 14BS314).

\end{document}